\documentclass[showpacs,aps,twocolumn]{revtex4}

\usepackage{float}
\usepackage{bm}
\usepackage{amsmath}
\usepackage{graphicx}
\usepackage{subfigure}
\usepackage[usenames,dvipsnames]{color}
\definecolor{darkblue}{RGB}{0,0,196}
\usepackage[colorlinks=true,linkcolor=darkblue,citecolor=darkblue,urlcolor=darkblue]{hyperref}

\usepackage{setspace}
\usepackage{footmisc}
\usepackage[makeroom]{cancel}
\usepackage{comment}
\usepackage{lineno}

\def\be{\begin{equation}}
\def\ee{\end{equation}}
\def\ba{\begin{eqnarray}}
\def\ea{\end{eqnarray}}

\usepackage{graphicx}
\usepackage{amsmath,bbm}
\usepackage{amssymb,bm}
\usepackage{yfonts}

\begin{document}
\title{Dissipative Properties and Isothermal Compressibility of Hot and Dense Hadron Gas using Non-extensive Statistics}
\author{Swatantra~Kumar~Tiwari}
\author{Sushanta~Tripathy}
\author{Raghunath~Sahoo\footnote{Corresponding author: $Raghunath.Sahoo@cern.ch$}}
\affiliation{Discipline of Physics, School of Basic Sciences, Indian Institute of Technology Indore, Simrol, Indore- 453552, INDIA}
\author{Nilotpal~Kakati}
\affiliation{Department of Physical Sciences, Indian Institute of Science Education and Research Mohali, Mohali-140306, INDIA}

\begin{abstract}
\noindent
We evaluate the transport properties such as shear viscosity ($\eta$), bulk viscosity ($\zeta$) and their ratios over entropy density ($s$) for hadronic matter using relativistic non-extensive Boltzmann transport equation (NBTE) in relaxation time approximation (RTA). In NBTE, we argue that the system far from equilibrium may not reach to an equilibrium described by extensive (Boltzmann-Gibbs (BG)) statistics but to a $q$-equilibrium defined by Tsallis non-extensive statistics after subsequent evolution, where $q$ denotes the degree of non-extensivity. We observe that $\eta/s$ and $\zeta/s$ decrease rapidly with temperature ($T$) for various $q$-values. As $q$ increases, the magnitudes of $\eta/s$ and $\zeta/s$ decrease with $T$. We also show the upper mass cutoff dependence of these ratios for a particular $q$ and find that they decrease with the increase in mass cutoff of hadrons. Further, we present the first estimation of isothermal compressibility ($\kappa_T$) using non-extensive Tsallis statistics at finite baryon chemical potential ($\mu_B$). It is observed that, $\kappa_T$ changes significantly with the degree of non-extensivity. We also study the squared speed of sound ($c_{s}^{2}$) as a function of temperature at finite baryon chemical potential for various $q$ and upper mass cutoffs. It is noticed that there is a strong impact of $q$ and mass cutoff on the behaviour of $c_{s}^{2}$.     


\end{abstract}
\date{\today}
\maketitle 
\section{Introduction}
\label{intro}
High-energy heavy-ion collisions provide an opportunity to probe the nuclear matter under extreme conditions $i. e.$ at high temperature and/or energy density. The space-time evolution of a system formed in heavy-ion collisions is governed by its dissipative properties such as shear viscosity ($\eta$), bulk viscosity ($\zeta$) etc. The spatial anisotropy created in heavy-ion collisions gets converted to the momentum anisotropy of the produced particles due to the pressure gradient. The equilibration of momentum anisotropy is governed by the shear viscosity. The AdS/CFT calculation shows that there is a lower bound to the value of shear viscosity to entropy density ratio ($\eta/s$) for any fluid found in nature. The lower bound (known as the Kovtun-Son-Starinets (KSS) bound) has been set to the value of $1/4\pi$ in natural units~\cite{Kovtun:2004de,Biro:2011bq}. The elliptic flow measurements at Relativistic Heavy-Ion Collider (RHIC) experiment have found that the $\eta/s$ is close to the KSS bound, which developed intense interest in this ratio of the strongly interacting matter described by quantum chromodynamics (QCD)~\cite{Romatschke:2007mq,Hirano:2005wx}. Also, a peak in Bulk viscosity to entropy density ratio ($\zeta/s$) is expected near QCD critical temperature ($T_{c}$), where conformal symmetry breaking might be significant as expected by various effective models~\cite{Dobado:2012zf,Sasaki:2008fg,Shushpanov:1998ce}. Other interesting thermodynamic properties are the isothermal compressibility ($\kappa_T$) and speed of sound ($c_s$), which are used to define the equation of state of the system and quantify the softest point of the phase transition along with the location of the critical point~\cite{Mukherjee:2017elm}. 

 Due to high multiplicities produced in high-energy collisions, the statistical models are more suitable to describe the particle production mechanism and to study the QCD thermodynamics. Such a statistical description of transverse momentum ($p_T$) of final state particles produced in high-energy collisions has been proposed to follow a thermalized Boltzmann-Gibbs (BG) distribution. But, a finite degree of deviation from the equilibrium statistical description of $p_T$ spectra has been observed by experiments at RHIC~\cite{star-prc75,phenix-prc83} and Large Hadron Collider (LHC)~\cite{alice1,alice2,alice3,cms}. In addition, the matter produced in high energy collisions evolves rapidly in a non-homogeneous way. Hence, the spatial configuration of the matter becomes non-uniform and the global equilibrium is not established \cite{Randrup:2009gp,Palhares:2009tf,Skokov:2008zp,Skokov:2009yu}. As a consequence of which some of the observables become non-extensive and develop a power-law tail in the spectra instead of exponential distributions. This also happens when there is local temperature fluctuation and long-range correlations in the produced system \cite{Wilk:1999dr}. In these cases, the use of BG distribution is questionable \cite{Wilk:2008ue,Wilk:2012zn,Wilk:2015pva,Wilk:2015hla,Wilk:2014zka}. Systems, where the usual ergodicity is violated, such as, the states in large systems involving long-range forces between particles and metastable states in small systems where the number of particles are relatively smaller, a generalized BG entropy has been introduced by C. Tsallis~\cite{Tsallis:1987eu,Tsallis:1999nq}. Recently, a growing attention has been paid towards the possible non-extensive effects in thermodynamics and statistical mechanics~\cite{Tsallis:2008mc,Kaniadakis} as well as to explain the particle spectra in high-energy hadronic and heavy-ion collisions~\cite{Thakur:2016boy,Sett:2015lja,Bhattacharyya:2015hya,Zheng:2015gaa,Tang:2008ud,De:2014dna}. The nuclear modification factor \cite{Tripathy:2017kwb,Tripathy:2016hlg} and elliptic flow of identified particles \cite{Tripathy:2017nmo} have also been successfully described by using Tsallis non-extensive statistics in Boltzmann Transport equation (BTE) with relaxation time approximation (RTA). 
 
 The aim of the present paper is to study various dissipative properties such as shear and bulk viscosities using the relativistic non-extensive  Boltzmann transport equation (NBTE), where we employ the RTA for the collision term. In NBTE, we assume that a non-equilibrium system, which dissipates energy and produces entropy, measured by the non-extensive parameter $q$, relaxes to a local $q$-equilibrium after a certain relaxation time $\tau$. The hadrons are treated as classical Boltzmann particles and the elastic scattering processes are taken into account. A similar approach has been employed in Ref. \cite{ref1}, where an extensive statistics is used in BTE to calculate the transport coefficients. Here, we extend the approach using Tsallis non-extensive statistics and calculate shear viscosity, bulk viscosity and their ratios over entropy density. First time, we quantify the isothermal compressibility for a hadronic matter using non-extensive statistics. We also calculate the squared speed of sound of hadron resonance gas at finite baryon chemical potential. In view of high-multiplicity events at the LHC and the observation of increase of particle multiplicity driving the system towards thermodynamic equilibrium \cite{Khuntia:2017ite}, it becomes important to study these dissipative properties of the systems which are away from equilibrium.

The paper runs as follows. In section~\ref{formulation}, the dissipative properties such as shear and bulk viscosities are derived using the relaxation time approximation of the relativistic NBTE. The formulation of isothermal compressibility and squared speed of sound in non-extensive statistics are also given. In section~\ref{result}, the results and discussions are presented. Finally, we summarize with the findings of this work in section~\ref{summary}.
 
\section{Formulation}
\label{formulation}
We follow the approach mentioned in Ref.~\cite{ref1} to calculate the dissipative properties such as shear and bulk viscosity for a hadronic matter using non-extensive statistics. We start with the BTE given by,
\begin{equation}
\frac{\partial f_p}{\partial t} + v_p^i \frac{\partial f_p}{\partial x^i} + F_{p}^i \frac{\partial f_p}{\partial p^i}= I(f_p),
\label{eq1}
\end{equation}
where $v_p^i$ is the velocity of $i^{\rm{th}}$ particle and $F_p^i$ is an external force acting on $i^{\rm{th}}$ particle. $I(f_p)$ is the collision integral which gives the rate of change of the non-equilibrium distribution function $f_p$ when the system approaches $q$-equilibrium. 

Assuming no external force and proceeding with the RTA, the collision integral can be approximated as,
\begin{equation}
I(f_p) \simeq -\frac{(f_p - f_p^0)}{\tau(E_p)},
\label{eq2}
\end{equation}
where $\tau(E_p)$ is the relaxation time or collision time. We take non-extensive Tsallis distribution as $f_p^0$~\cite{Bezerra:2002gi} near the local rest frame of the fluid, where the system is described locally by $T$, $\mu_B$ and fluid velocity, $\bf u$, which change slowly in space and time~\cite{Gavin:1985ph}. The thermodynamically consistent Tsallis distribution, ($f_p^0$)~\cite{ref5} in the Boltzmann's approximation is given as,
\begin{equation}
f_p^0 = \frac{1}{\Big[1 + (q-1)\Big(\displaystyle \frac{E_p - \bf{p.u} - \mu}{T}\Big)\Big]^{\displaystyle \frac{q}{q-1}}},
\label{eq3}
\end{equation}
where $\bf{u}$ is the fluid velocity. $T$ and $\mu$ are temperature and chemical potential, respectively. $\mu = b\mu_B + s\mu_s$, where $b$ and $s$ are baryon and strangeness quantum numbers, respectively. $\mu_B$ and $\mu_s$ are baryon and strange chemical potentials. For the sake of simplicity, we do not consider the strangeness neutrality condition.  

Now, the stress-energy tensor ($T^{\mu\nu}$) can be written as,
\begin{equation}
T^{\mu\nu} = T^{\mu\nu}_{0} +T^{\mu\nu}_{dissi},
\label{eq4}
\end{equation}
where $T^{\mu\nu}_{0}$ is the ideal part and $T^{\mu\nu}_{dissi}$ is the dissipative part of the stress-energy tensor. In the hydrodynamical description of QCD, shear and bulk viscosities enter in the dissipative part of the stress-energy tensor, which can be written (in the local Lorentz frame) as~\cite{Gavin:1985ph},
\begin{equation}
T_{dissi}^{ij} = -\eta \left(\frac{\partial u^i}{\partial x^j} + \frac{\partial u^j}{\partial x^i}\right) - \left(\zeta - \frac{2}{3}\eta\right) \frac{\partial u^i}{\partial x^j} \delta^{ij}.
\label{eq5}
\end{equation} 
In terms of distribution function, this can be expressed as,
\begin{equation}
T_{dissi}^{ij} = \int \frac{d^3p}{(2\pi)^3}\frac{p^ip^j}{E_p}\delta f_p,
\label{eq6}
\end{equation}
where $\delta f_p$ is the deviation of the distribution function from the $q$-equilibrium and is given by (from Eqs.~\ref{eq1} and~\ref{eq2}),
\begin{equation}
\delta f_p = -\tau (E_p) \left(\frac{\partial f_p^0}{\partial t} + v_{p}^i \frac{\partial f_p^0}{\partial x^i}\right).
\label{eq7}
\end{equation}
Assuming a steady flow of the form $u^i = (u_x(y), 0, 0)$ and space-time independent temperature, Eq.~\ref{eq5} simplifies to $T^{xy} = -\eta \partial u_x / \partial y$. Now, from Eqs.~\ref{eq6} and~\ref{eq7}, we get (using $\mu = 0$),

\begin{equation}
T^{xy} = \left\lbrace - \frac{1}{T} \int \frac{d^3p}{(2\pi)^3} \tau(E_p) \left(\frac{p_xp_y}{E_p}\right)^2 q (f_p^0)^{\frac{(2q-1)}{q}}\right\rbrace \frac{\partial u_x}{\partial y}.
\label{eq9}
\end{equation}
Thus the coefficient of shear viscosity for a single component of hadronic matter can be expressed as,
\begin{equation}
\eta = \frac{1}{15T} \int \frac{d^3p}{(2\pi)^3} \tau(E_p) \frac{p^4}{E_p^2}q(f_p^0)^{\frac{(2q-1)}{q}}.
\label{eq10}
\end{equation}

The bulk viscosity ($\zeta$) is related to the dissipation when the system is uniformly compressed. We take the trace of Eq.~\ref{eq5} and get,
\begin{equation}
(T_{dissi})^{i}_{i} = -3\zeta \frac{\partial u^{i}}{\partial x^{i}}.
\label{eq11}
\end{equation}
Using Eq.~\ref{eq6}, the above equation modifies as,
\begin{equation}
(T_{dissi})^{i}_{i} = \int \frac{d^3p}{(2\pi)^3} \frac{p^2}{E_p} \delta f_p.
\label{eq12}
\end{equation}
Now, using the conservation law for energy-momentum $i.e.$ $\partial_{\mu} T^{\mu\nu} = 0$, one can obtain~\cite{Gavin:1985ph},
\begin{equation}
\frac{\partial \varepsilon}{\partial t}=-w\nabla.u,
\label{eq13}
\end{equation}
and
\begin{equation}
w\frac{\partial u}{\partial t}=-\nabla P,
\label{eq14}
\end{equation}
where, $w=\varepsilon + P$ is the enthalpy density, $\varepsilon$ is the energy density and $P$ is the pressure. Here, we cannot take steady flow. Thus, Eq.~\ref{eq7} takes the form,
\begin{equation}
\delta f_p = -\tau(E_p) \frac{q}{T}(f_p^0)^{\frac{(2q-1)}{q}}\Big[E_p\frac{w}{C_VT}\nabla.u - (v.\nabla)(\bf p.u)\Big],
\label{eq15}
\end{equation}
where $C_V$ is the heat capacity at constant volume. Using Eqs~\ref{eq12} and~\ref{eq15} and manipulating with the help of conservation laws, the expression for $\zeta$ is written as~\cite{Gavin:1985ph}, 
\begin{equation}
\zeta = \frac{1}{T} \int \frac{d^3p}{(2\pi)^3} \tau(E_p) q(f_p^0)^{\frac{(2q-1)}{q}}\left(\frac{E_pw}{C_VT} - \frac{E_p}{3} + \frac{m^2}{3E_p}\right)^2.
\label{eq16}
\end{equation}
Now, using $dP=sdT=wdT/T$, where $s$ is the entropy density, the above equation is reduced to,
\begin{equation}
\zeta = \frac{1}{T} \int \frac{d^3p}{(2\pi)^3} \tau(E_p)q(f_p^0)^{\frac{(2q-1)}{q}} \left(E_p c^2_{s} - \frac{p^2}{3E_p}\right)^2.
\label{eq17}
\end{equation}
Here, $c^2_{s}=s/C_V$ is the squared speed of sound at constant baryon density. 
For a multi-component hadron gas at finite chemical potential, the shear and bulk viscosities are written as,
\begin{equation}
\eta = \frac{1}{15T} \sum_{a} \int \frac{d^3p}{(2\pi)^3} \frac{p^4}{E_a^2}q \left(\tau_a (f_a^0)^{\frac{(2q-1)}{q}} + \bar{\tau_a} (\bar{f_a^0})^{\frac{(2q-1)}{q}}\right),
\label{eq19}
\end{equation}
and,
\begin{equation}
\begin{split}
\zeta =  \frac{1}{T}\sum_{a}\int \frac{d^3p}{(2\pi)^3} \tau_a q (f_a^0)^{\frac{(2q-1)}{q}} \left[ E_a c^2_{s}+ \left(\frac{\partial P}{\partial n_B}\right)_{\epsilon}- \frac{p^2}{3E_a}\right]^2 \\ 
 + \frac{1}{T}\sum_{a}\int \frac{d^3p}{(2\pi)^3}\bar{\tau_a}q (\bar{f_a^0})^{\frac{(2q-1)}{q}} \left[E_a c^2_{s} - \left(\frac{\partial P}{\partial n_B}\right)_{\epsilon}- \frac{p^2}{3E_a}\right]^2.
\label{eq20}
\end{split}
\end{equation}
Here $E_a^2 = p^2 + m_a^2$ and the barred quantities represent the antiparticles. $f_a^0$ is the distribution function for $a^{\rm{th}}$ particle. Now, the energy dependent relaxation time is given as, 
\begin{equation}
\tau^{-1} (E_a) = \sum_{bcd} \int \frac{d^3p_b d^3p_c d^3p_d}{(2\pi)^3 (2\pi)^3 (2\pi)^3} W(a,b \rightarrow c,d) f_b^0,
\label{eq21}
\end{equation}
where $W(a,b \rightarrow c,d)$ is the transition rate defined as,
\begin{equation}
W(a,b \rightarrow c,d) = \frac{{2\pi}^4\delta(p_a+p_b-p_c-p_d)}{2E_a2E_b2E_c2E_d} |\mathcal{M}|^2.
\label{eq22}
\end{equation}
Here $|\mathcal{M}|$ is the transition amplitude. In the center-of-mass frame, Eq.~\ref{eq21} can be simplified as,
\begin{equation}
\begin{split}
\tau^{-1}(E_a) &= \sum_{b} \int \frac{d^3p_b}{(2\pi)^3} \sigma_{ab} \frac{\sqrt{s - 4m^2}}{2E_a2E_b} f_b^0 \\
&\equiv \sum_{b} \int \frac{d^3p_b}{(2\pi)^3} \sigma_{ab} v_{ab} f_b^0,
\end{split}
\label{eq23}
\end{equation}
where $v_{ab}$ is the relative velocity and $\sqrt{s}$ is the center-of-mass energy. $\sigma_{ab}$ is the total scattering cross-section in the process $a(p_a) + b(p_b) \rightarrow a(p_c) + b(p_d)$. For further simplification, $\tau(E_a)$ can be approximated to averaged relaxation time ($\widetilde{\tau}$)~\cite{Moroz:2013haa} and it can be obtained from Eq.~\ref{eq23} by averaging over $f_a^0$ as,

\begin{equation}
\begin{split}
\widetilde{\tau_a}^{-1}(E_a) &= \frac{\int \displaystyle \frac{d^3p_a}{(2\pi)^3}\tau^{-1}(E_a) f_a^0}{\int \displaystyle \frac{d^3p_a}{(2\pi)^3}f_a^0} \\
&= \sum_{b} \displaystyle \frac{\int \displaystyle \frac{d^3p_a}{(2\pi)^3} \displaystyle \frac{d^3p_b}{(2\pi)^3} \sigma_{ab}v_{ab}f_a^0f_b^0}{\int \displaystyle \frac{d^3p_a}{(2\pi)^3}f_a^0}  \\
 &= \sum_{b} n_b \langle \sigma_{ab} v_{ab}\rangle, 
\end{split}
\label{eq24}
\end{equation}
here $\displaystyle n_b = \int \frac{d^3p_b}{(2\pi)^3}f_b^0$ is the number density of $b^{th}$ hadronic species. The thermal average for the scattering of same species of particles with constant cross-section at zero baryon density can be calculated as follows~\cite{ref1,ref2, ref3}. 
\begin{equation}
\begin{split}
\left\langle \sigma_{ab}v_{ab} \right \rangle = \frac{\sigma\int d^3p_a d^3p_b v_{ab}e_{q}^{-E_{a}/T}e_{q}^{-E_{b}/T}}{\int d^3p_a d^3p_be_{q}^{-E_{a}/T}e_{q}^{-E_{b}/T}}.
\end{split}
\label{eq25}
\end{equation}
Here $e_q^{(x)}$ is the $q$-exponential which is defined as $e_q^{(x)} = [1+(q-1)x]^{q/(q-1)}$.
The momentum space volume elements can be written as, 
\begin{equation}
\begin{split}
d^3p_a d^3p_b= 8\pi^2 p_a p_b dE_a dE_b d\cos\theta.
\end{split}
\label{eq26}
\end{equation}

The numerator in Eq.~\ref{eq25} is written as,
\begin{widetext}
\begin{equation}
\sigma \int d^3p_a d^3p_b v_{ab}e_{q}^{-E_{a}/T}e_{q}^{-E_{b}/T}= \sigma \int 8\pi^2 p_a p_b dE_a dE_b d\cos \theta~e_{q}^{-E_{a}/T}e_{q}^{-E_{b}/T}
\times \frac{\sqrt{(E_aE_b-p_ap_b\cos\theta)^2-(m_am_b)^2}}{E_aE_b-p_ap_b\cos\theta}.
\label{eq27}
\end{equation}
\end{widetext}
and the denominator is written as,
\begin{eqnarray}
\int d^3p_a d^3p_b e_{q}^{-E_{a}/T}e_{q}^{-E_{b}/T}= \int 8\pi^2 p_a p_b dE_a dE_b  \nonumber\\
                                                                   \times d\cos\theta e_{q}^{-E_{a}/T}e_{q}^{-E_{b}/T}.
\label{eq28}
\end{eqnarray}

Now, using the generalized Tsallis non-extensive statistics for $q$-equilibrium, the $\left\langle \sigma_{ab}v_{ab}\right\rangle$ takes the following form,
\begin{widetext}
\begin{equation}
\left\langle \sigma_{ab}v_{ab} \right \rangle =  \frac{ \sigma \int 8\pi^2 p_a p_b dE_a dE_b d\cos\theta~e_{q}^{-E_{a}/T}e_{q}^{-E_{b}/T}\times \frac{\sqrt{(E_aE_b-p_ap_b\cos\theta)^2-(m_am_b)^2}}{E_aE_b-p_ap_b\cos\theta}}{\int 8\pi^2 p_a p_b dE_a dE_b d\cos\theta~e_{q}^{-E_{a}/T}e_{q}^{-E_{b}/T}}.
\label{eq29}
\end{equation}
\end{widetext}
Here $\sigma$ is the cross-section used as a parameter in the calculations. $E_a$ and $E_b$ are integrated in the limit $m_a$ to $\infty$ and $m_b$ to $\infty$, respectively. The limit of integration for $\cos\theta$ is -1 to 1. The relaxation time is calculated by using Eqs.~\ref{eq24} and~\ref{eq29}. The other thermodynamical quantities in non-extensive statistics are calculated as~\cite{ref5},
\begin{alignat}{3}
& n = g \int \frac{d^3p}{(2\pi)^3} \left[1 + (q-1) \frac{E-\mu}{T}\right] ^ {-\frac{q}{q-1}} \label{eq30} \\
& \epsilon = g \int \frac{d^3p}{(2\pi)^3} E \left[1 + (q-1)\frac{E-\mu}{T}\right] ^ {-\frac{q}{q-1}} \label{eq31} \\
& P= g \int \frac{d^3p}{(2\pi)^3} \frac{p^2}{3E} \left[1 + (q-1)\frac{E-\mu}{T}\right] ^ {-\frac{q}{q-1}}.
\label{eq32}
\end{alignat}
$n$, $\epsilon$ and $P$ are the number density, energy density and pressure of hadrons, respectively. The non-extensive entropy density, $s$ can be calculated from the above expression as,
\begin{equation}
s= \frac{\epsilon + P-\mu n}{T}.
\label{eq33}
\end{equation}
We use the basic expression for addition of entropies in non-extensive statistics while calculating for the multi-component hadron gas, which is given by,
\begin{equation}
s(A+B)= s(A)+s(B)-(q-1)s(A)s(B),
\label{eq34}
\end{equation}
where $s(A+B)$ is the total entropy of $A$ and $B$. $s(A)$ and $s(B)$ are the entropies of $A$ and $B$, respectively. 

We have also calculated the isothermal compressibility for hadron gas using non-extensive statistics. The isothermal compressibility ($\kappa_T$) is defined as \cite{ref6},
\begin{equation}
\kappa_T = - \frac{1}{V} \frac{\partial V}{\partial P},
\label{eq35}
\end{equation}
where $V$ is the volume of the system. Again in terms of fluctuation and average number, isothermal compressibility can be expressed as \cite{ref6,ref7},

\begin{equation}
\left\langle (N - \left\langle N\right\rangle)^2 \right\rangle = var(N) = \frac{T \left\langle N \right\rangle ^2}{V} \kappa_T.
\label{eq36}
\end{equation}

Since we use the thermodynamically consistent Tsallis statistics, the above thermodynamical relation is valid in this case. Using the basic thermodynamical relation $\displaystyle \left\langle (N - \left\langle N \right\rangle)^2\right\rangle  = VT \frac{\partial n}{\partial \mu}$, Eq.~\ref{eq36} can further be expressed in terms of number density and compressibility as, 
\begin{equation}
\frac{1}{\kappa_T} =  \sum_a \displaystyle \frac{n_{aq}^2}{\left(\displaystyle \frac{\partial n_{aq}}{\partial \mu} \right)},
\label{eq37}
\end{equation}
where $\displaystyle \frac{\partial n_q}{\partial \mu}$ is given as,
\begin{equation}
\frac{\partial n_q}{\partial \mu} = \frac{gq}{T} \int \frac{d^3p}{(2\pi)^3} \left[1 + (q-1)\frac{E - \mu}{T}\right] ^ \frac{1-2q}{q-1}.
\label{eq38}
\end{equation}

In hydrodynamics, speed of sound plays an important role in understanding the Equation of State (EoS) and hence, the associated phase transition. This is because of the fact that speed of sound depends on the properties of the medium. The squared speed of sound ($c_s^2$) is defined as,
\begin{equation}
c_s^2 = \Big(\frac{\partial P}{\partial \epsilon}\Big)_{s/n}.
\label{eq39}
\end{equation}
The expression of $c_s^2$ at finite baryon chemical potential is written as~\cite{ref8}, 
\begin{equation}
c_s^2 (T, \mu_B) = \frac{\left(\frac{\partial P}{\partial T}\right) + \left(\frac{\partial P}{\partial \mu_B}\right) \left(\frac{d \mu_B}{dT}\right)}{\left(\frac{\partial \epsilon}{\partial T}\right) + \left(\frac{\partial \epsilon}{\partial \mu_B}\right)\left(\frac{d \mu_B}{dT}\right)}.
\label{eq40}
\end{equation}
The term $\frac{d \mu_B}{dT}$ can be calculated using the constant entropy per particle ($s/n$) condition~\cite{ref8,Tiwari:2011km}, and given as,
\begin{equation}
\frac{d \mu_B}{dT} = \frac{s \left(\frac{\partial n}{\partial T}\right) - n \left(\frac{\partial s}{\partial T}\right)}
{n \left(\frac{\partial s}{\partial \mu_B}\right) - s \left(\frac{\partial n}{\partial \mu_B}\right)}.
\label{eq41}
\end{equation}

\section{Results and Discussions}
\label{result}

\begin{figure}[h]
\centering
\includegraphics[width = 250px, height = 180px]{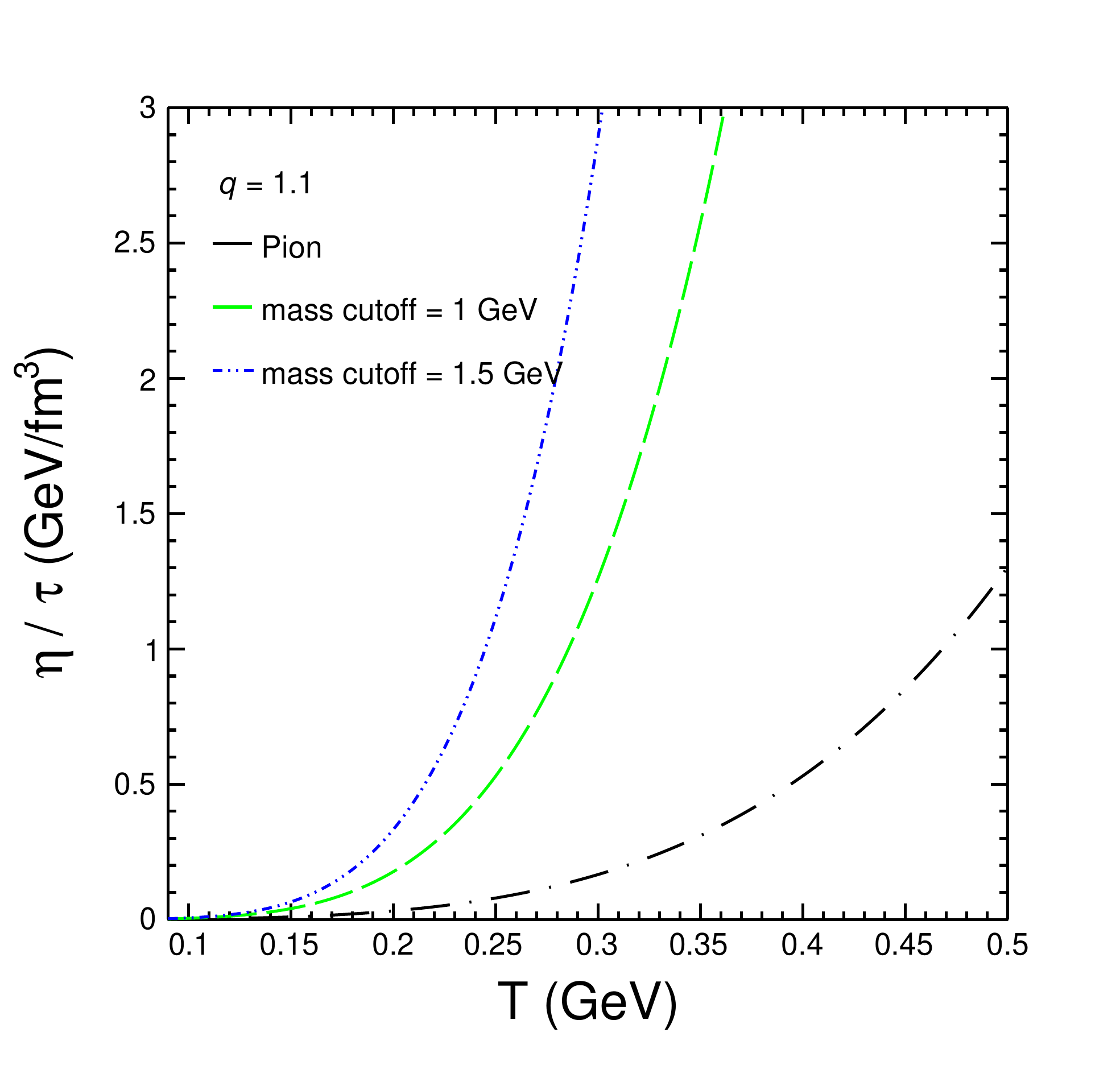}
\caption{(Color online) $\eta/\tau$ versus temperature for various mass cutoff at zero baryon chemical potential.}
 \label{eta_tau_m} 
\end{figure}

\begin{figure}[h]
\centering
\includegraphics[width = 250px, height = 180px]{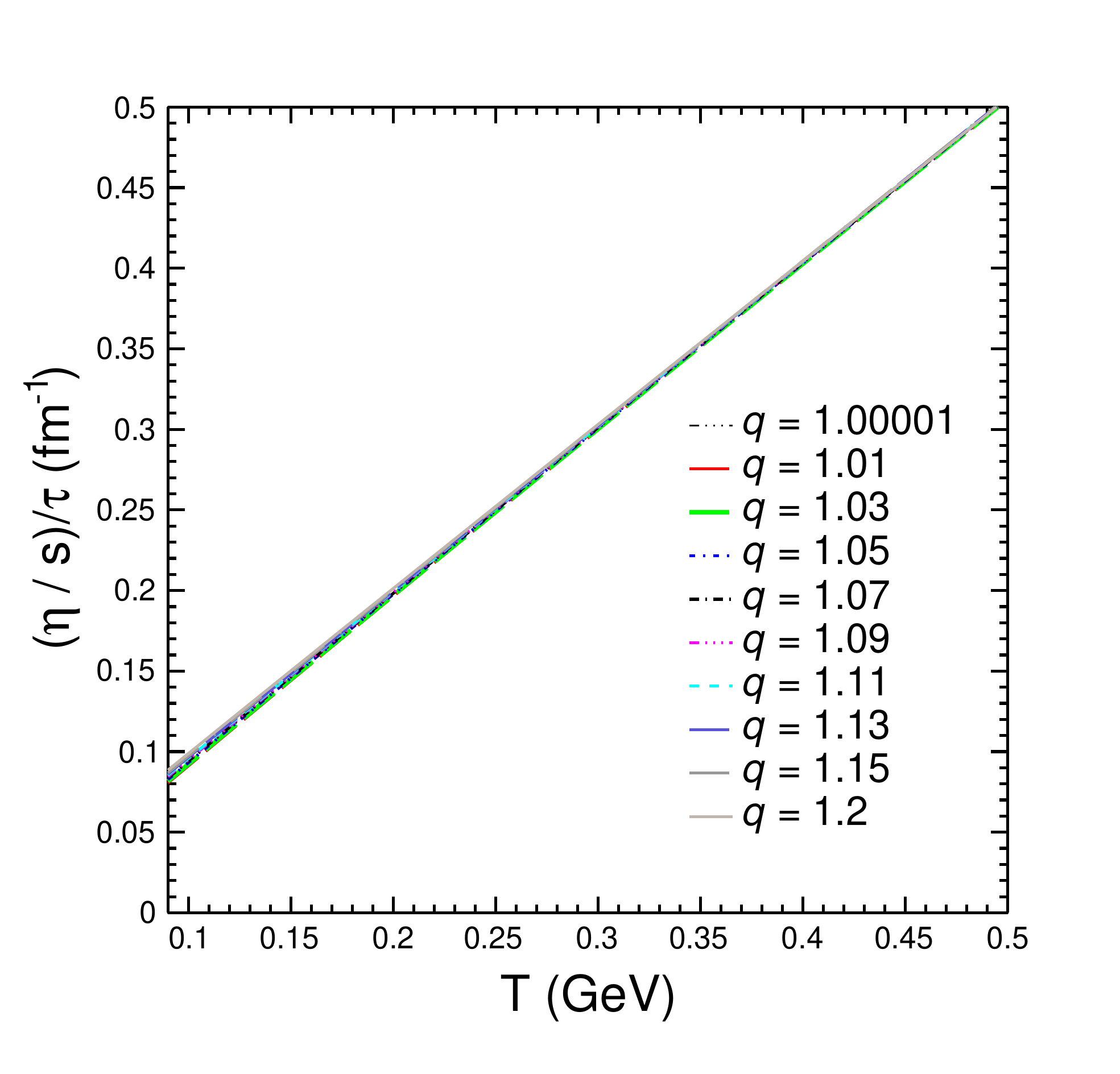}
\caption{(Color online) $(\eta/s)/\tau$ versus temperature for various $q$- values at zero baryon chemical potential.}
 \label{eta_tau_q} 
\end{figure}

In this section, we present the results and discussions of the dissipative and thermodynamic properties of hadronic matter calculated in the NBTE using relaxation time approximation, where non-extensive statistics is used as the $q$-equilibrium distribution function. The non-extensive parameter, q represents the degree of non-extensivity or in other words the tendency of the system to stay away from the thermodynamic equilibrium. The relaxation time of the system is thus correlated with the q-parameter. The non-extensive parameter, q is related to local temperature fluctuation and possible long-range correlation in the system~\cite{Wilk:1999dr}, which is also responsible for the deviation of the system being away from the thermodynamic equilibrium. It has been reported that in the hadronic and elementary collisions where the spectral analysis shows a deviation of the system from the BG equilibrium, the q-value is 1.22~\cite{Beck:2000nz}. We study the response of the system as a function of q-parameter. The present study will be very important in view of the multiplicity dependent analysis done in $p+p$ collisions at the LHC. We have seen explicitly from the identified particle spectra in $p+p$ collisions that with higher multiplicity classes, the $q$-value goes towards one \cite{Khuntia:2017ite}. We consider all hadrons and their resonances with a mass cutoff of 2.0 GeV. First, we estimate various dissipative properties of hadron gas such as shear viscosity ($\eta$), bulk viscosity ($\zeta$) and their ratios to entropy density at zero baryon chemical potential ($\mu_B$ = 0). We use Eqs.~\ref{eq19} and~\ref{eq33} to estimate $\eta$ and $s$, respectively for zero baryon chemical potential. Here, we calculate the relaxation time ($\tau$) of $a^{\rm{th}}$ hadron by using Eq.~\ref{eq24}, where a constant hadronic collision cross-section ($\sigma$) = 11.3 mb is taken~\cite{ref1}. In order to check the validity of formulations discussed above, we have presented the calculation of $\eta/\tau$ for pion and various mass cutoffs at $q$ = 1.1 in Fig.~\ref{eta_tau_m}. We observe that $\eta/\tau$ initially varies very slowly with temperature and as temperature increases, it increases rapidly. The results for pion exactly matches with those of presented in Ref.~\cite{Biro:2011bq}, where the degeneracy ($g$) of pion is taken to be 1. As we add more hadrons and their resonances into the system, $\eta/\tau$ increases rapidly at a lower $T$. These findings suggest that the shear viscosity becomes large with the increasing number of particles into the system. We have also shown the variations of the ratio $(\eta/s)/\tau$ of pion with $T$ at various $q$- values in Fig.~\ref{eta_tau_q}. Again, we find the same behaviour as observed in Ref.~\cite{Biro:2011bq} $i. e.$ there is almost negligible effect of the non-extensive parameter on this ratio particularly at higher temperature. This is due to the fact that entropy density ($s$) also increases with the same amount as $\eta/\tau$ with $q$.    
\begin{figure}[htp]
\vspace{-0.4cm}
\begin{center}
 \includegraphics[width=7cm,height=7cm]{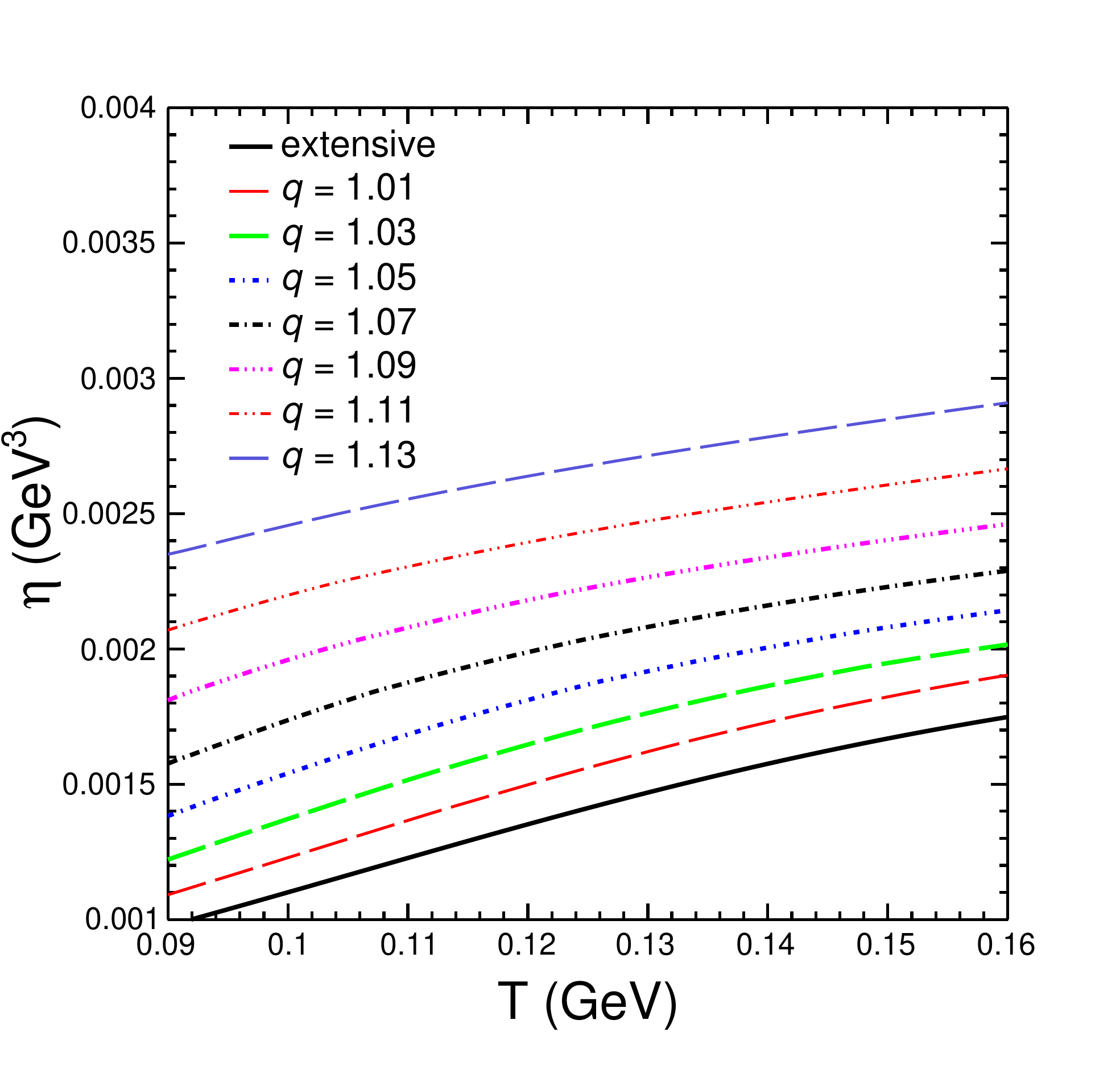}\\
 \includegraphics[width=7cm,height=7cm]{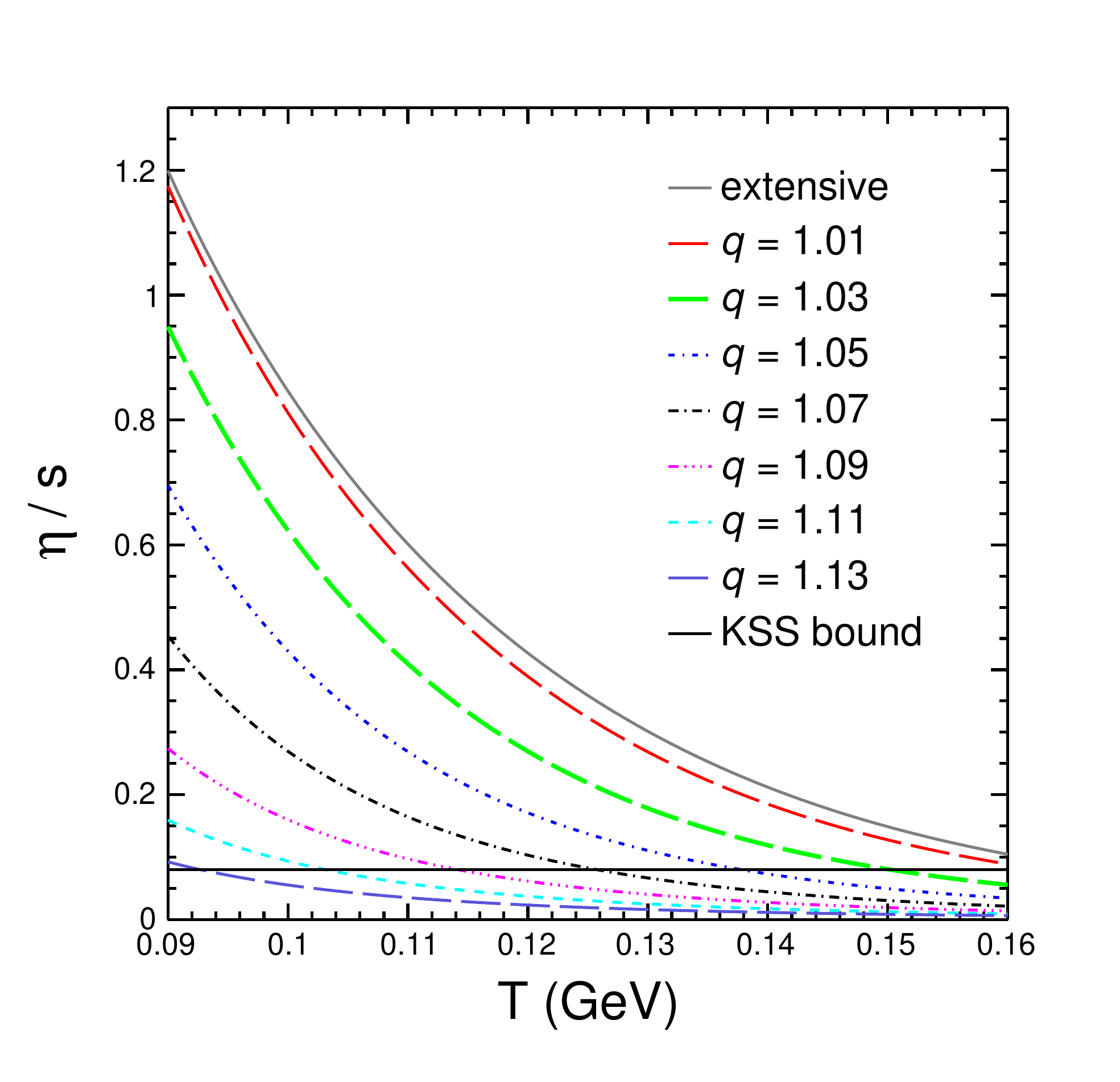}\\
 \includegraphics[width=7cm,height=7cm]{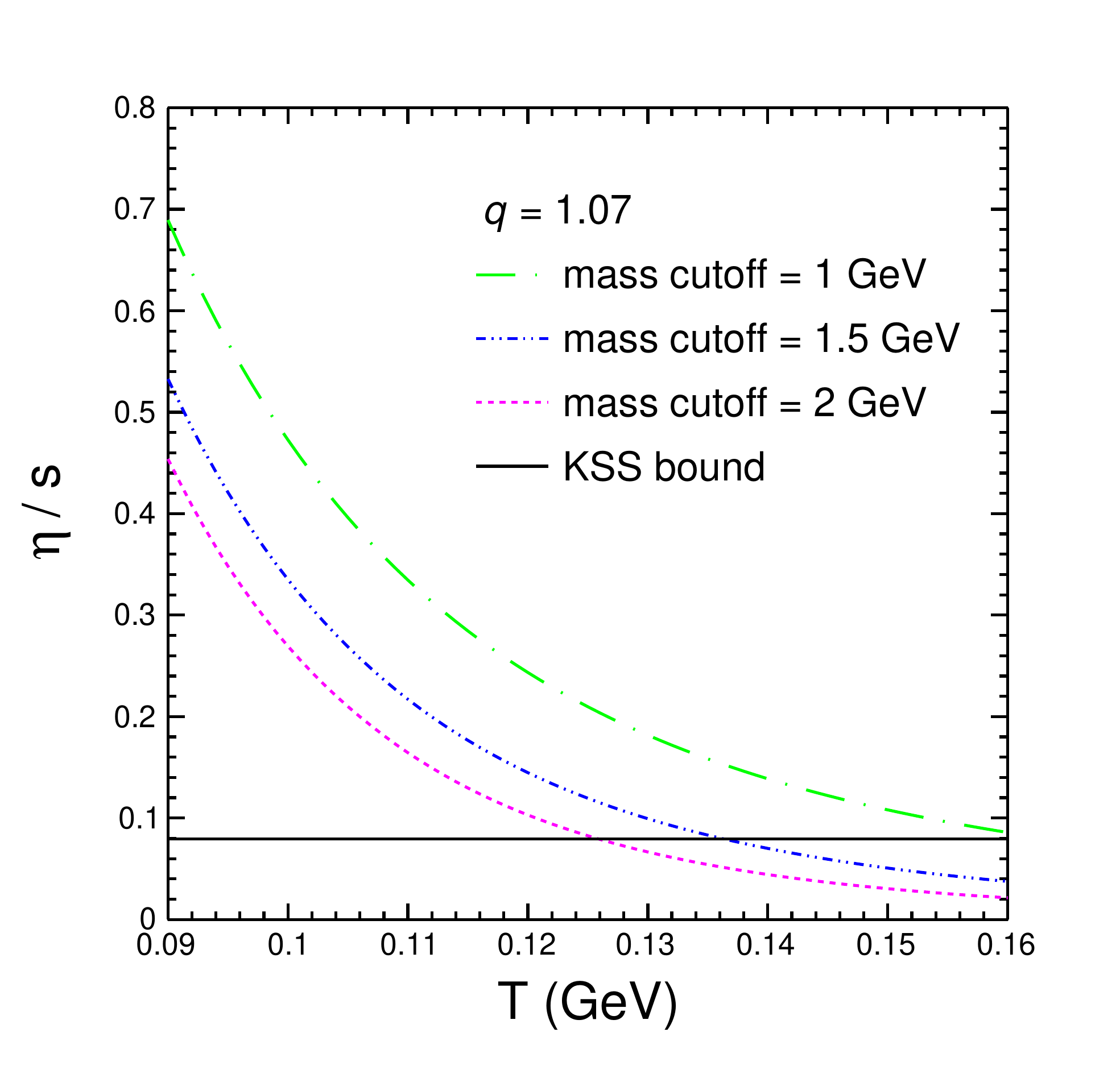} 
  \caption{(Color online) {\bf Top panel} shows the variations of shear viscosity ($\eta$) as a function of temperature for different $q$-values. {\bf Middle panel} shows $\eta / s$ versus temperature for different $q$-values. {\bf Lower panel} presents the variations of $\eta/s$ with respect to temperature at $q$ = 1.07 for various upper mass cutoff of hadron resonance gas.} 
\label{eta}
  \end{center}
 \end{figure}
 
The results of shear viscosity ($\eta$) of hadrons as a function of temperature ($T$) are presented in the top panel of Fig.~\ref{eta} for various $q$-values. We find that $\eta$ increases with temperature in a similar way as observed in Ref.~\cite{ref1}, where Excluded-Volume Model with extensive statistics is used. $\eta$ has smaller values for lower $q$, which goes in line with the observation that the values of the transport coefficients become smaller as the system goes towards equilibrium. Middle panel presents the variations of shear viscosity to entropy density ratio ($\eta/s$) with respect to temperature at zero baryon chemical potential, where different lines are for different $q$-values. Here, solid horizontal line is the KSS bound. The general behaviour of $\eta/s$ is similar to that observed in Ref.~\cite{ref1}. This ratio decreases with $T$ at all $q$-values due to rapid increase in entropy density. 

\begin{figure}[h]
\centering
\includegraphics[width = 250px, height = 180px]{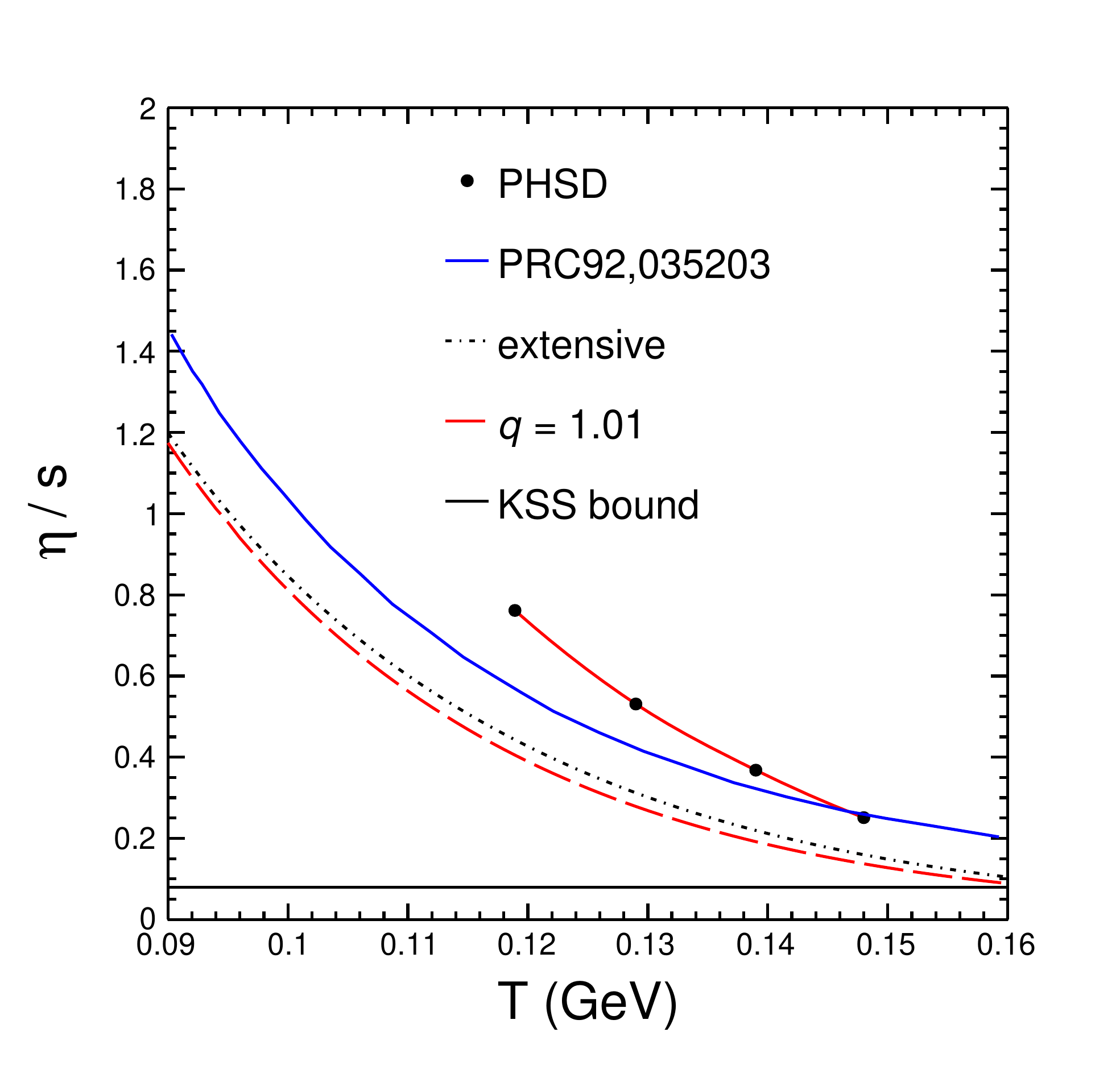}
\caption{(Color online) Comparison of $\eta$/s results estimated in various theoretical models.}
 \label{comparison} 
\end{figure}

As the degree of non-extensivity, $q$ increases, the magnitude of $\eta/s$ decreases and goes below the KSS bound at a higher $T$. The temperature at which $\eta/s$ goes below the KSS bound decreases with increasing $q$. This can be understood as the drastic change in entropy density of multicomponent hadron gas with $T$, which makes the value of the ratio, $\eta/s$ very small. This explanation has been given as a counterexample to the KSS bound in Ref.~\cite{Cohen:2007qr}. Also, in certain field theories~\cite{Buchel:2008vz,Rebhan:2011vd,Mamo:2012sy} KSS bound has been shown to be violated. This is a very important finding while understanding the dissipative properties, non-extensivity and the effect of temperature of the system. For comparison, we have also shown the calculation of $\eta/s$ for hadron gas at zero baryon chemical potential using Boltzmann-Gibbs distribution function. This is shown by the grey solid line in the middle panel of Fig.~\ref{eta}.
 $\eta/s$ as a function of temperature is presented in the lower panel of the figure at $q$ = 1.07 for various upper mass cutoff of hadrons. Study of the response of the system as a function of temperature for various mass cutoffs is a requirement at higher collision energies. With increase of collision energies heavier resonances are produced. We find that the magnitude of $\eta/s$ decreases with the increase of mass cutoff due to the contributions of heavier resonances in entropy density of the system.
\begin{figure}[htp]
\vspace{-0.4cm}
\begin{center}
 \includegraphics[width=7cm,height=7cm]{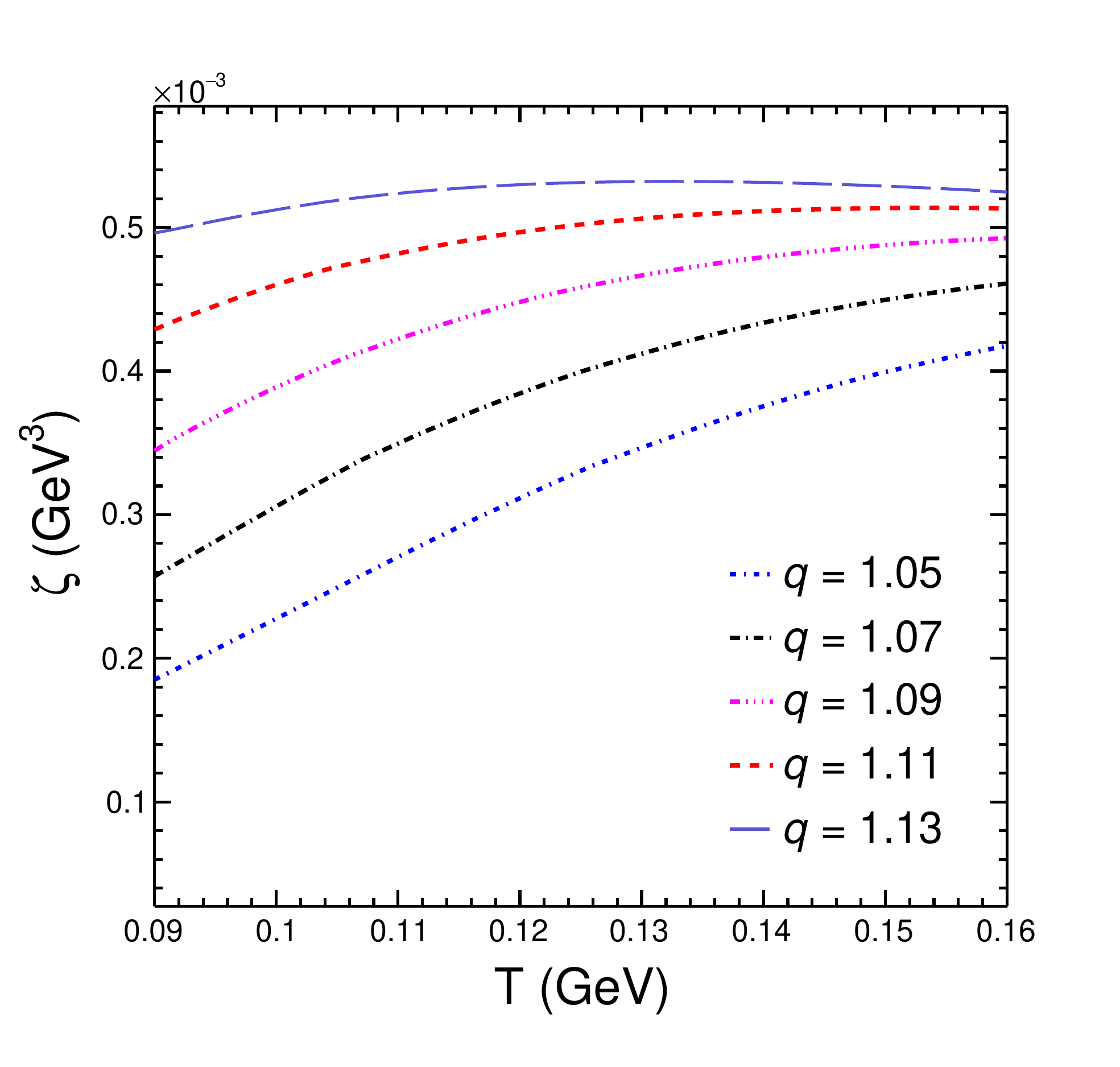}\\
 \includegraphics[width=7cm,height=7cm]{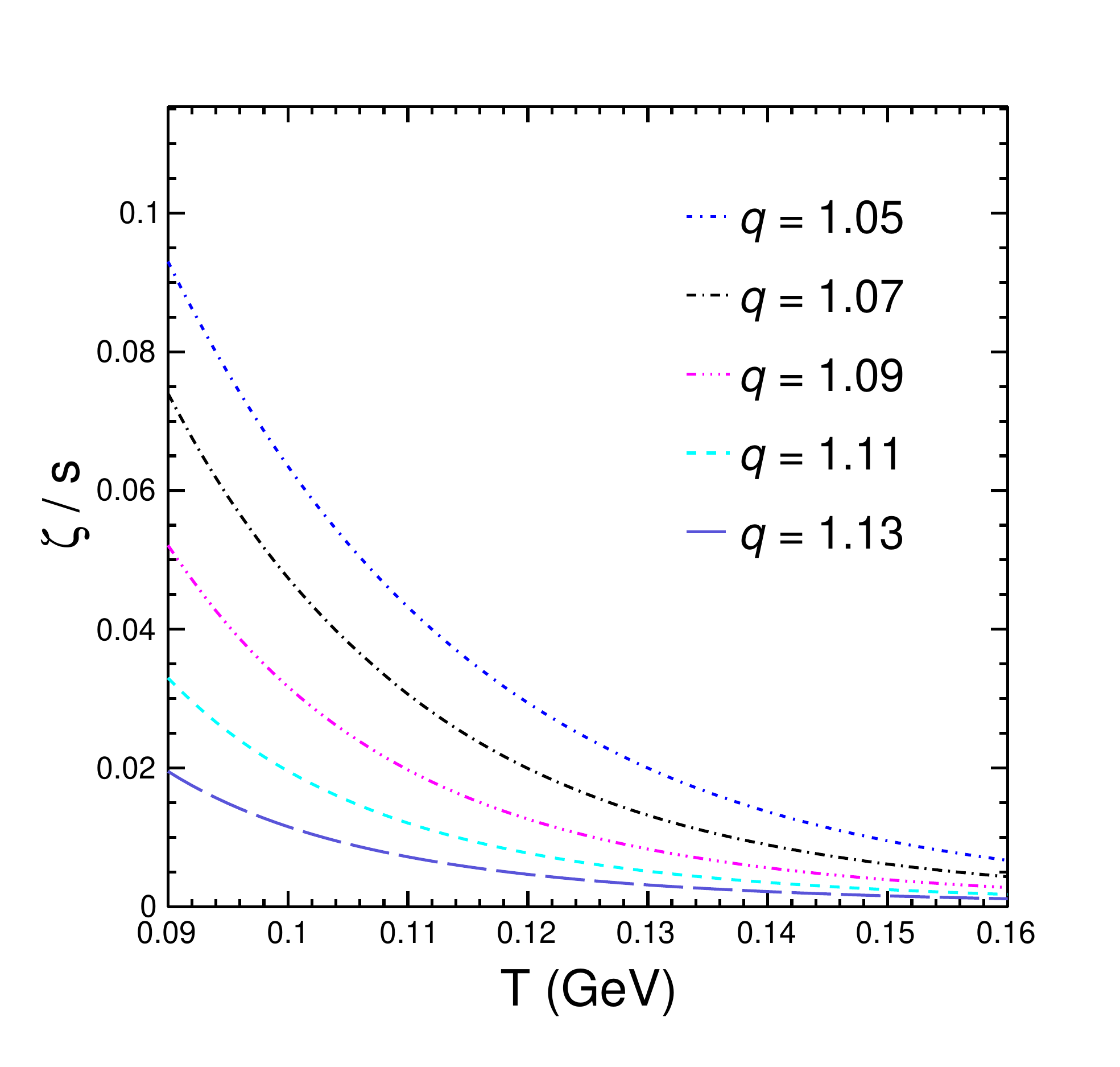}\\
 \includegraphics[width=7cm,height=7cm]{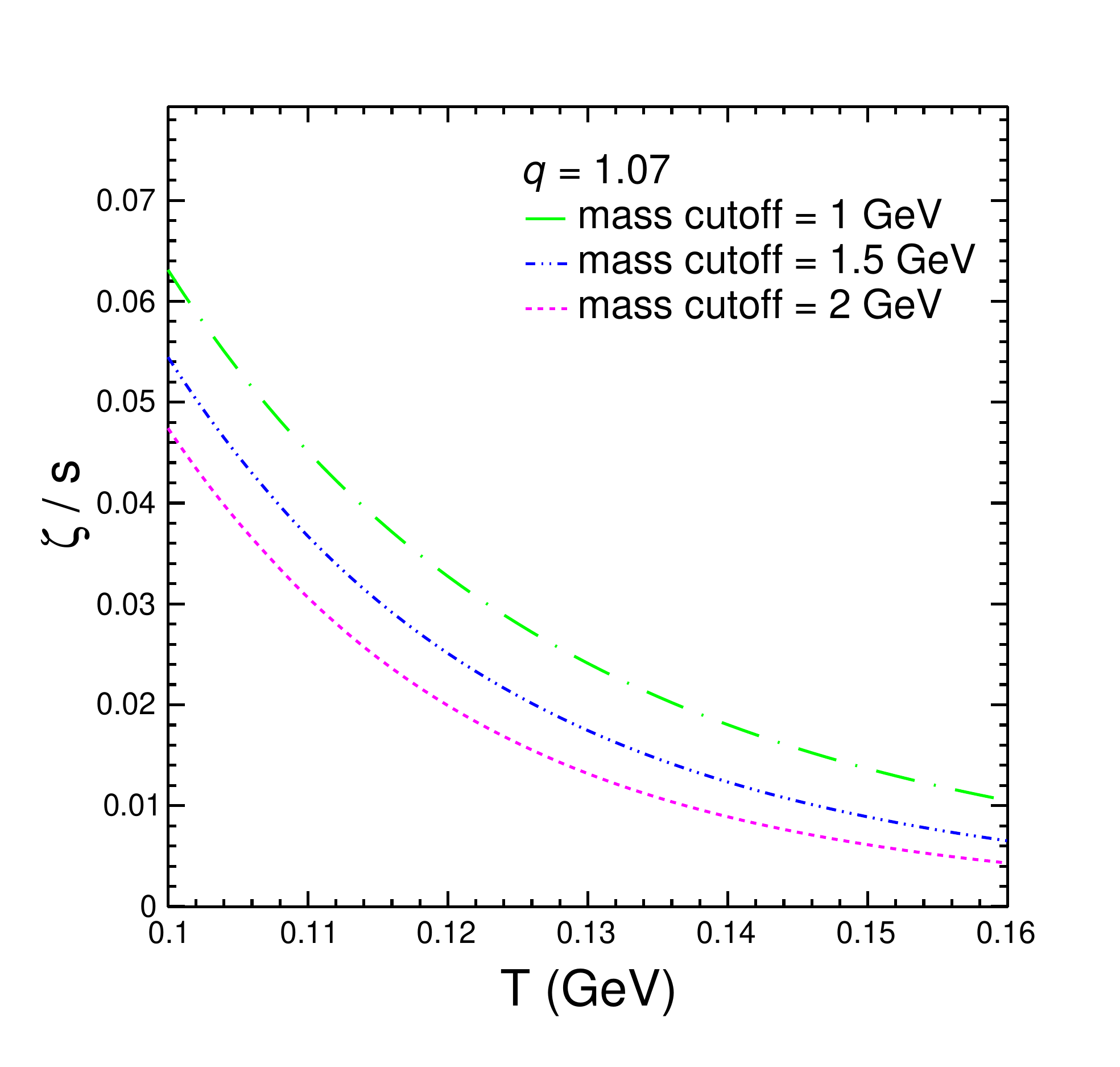} 
  \caption{(Color online) {\bf Top panel} represents bulk viscosity ($\zeta$) as a function of temperature for different $q$-values. {\bf Middle panel} shows $\zeta / s$ versus temperature for different $q$-values. {\bf Lower panel} presents the variations of $\zeta/s$ with respect to temperature at $q$ = 1.07 for various upper mass cutoff of hadron resonance gas.} 
\label{zeta}
  \end{center}
 \end{figure}

  In Fig.~\ref{comparison}, we have shown the comparison of $\eta$/s as a function of T obtained in different model calculations. Here, the red solid line is $\eta$/s for hadron gas calculated using parton-hadron-string dynamics (PHSD)~\cite{Ozvenchuk:2012kh}, where RTA with Boltzmann-Gibbs distribution as equilibrium distribution function is used. The blue line is the result taken from Ref.~\cite{ref1}, where BTE with RTA is used and BG distribution is taken as the solution of BTE. The red dotted and black dash-dotted lines are the estimations of the present work for $q$ = 1.01 and for extensive case ($q=1$), respectively. The difference between the PHSD and our results for $q$ = 1.01$ (\sim$1) is due to the difference in estimation of relaxation time. Although we have followed a similar procedure as done in Ref.~\cite{ref1} using non-extensive distribution function, for simplicity, we do not include the excluded-volume correction in the estimation of thermodynamical observables such as number density, entropy density etc., as is done in Ref.~\cite{ref1}. We also see from the figure that when q$\rightarrow$1, the results are closer to those obtained for extensive case.
 
In Fig.~\ref{zeta}, we discuss the results of bulk viscosity ($\zeta$) and its ratio to entropy density ($s$) calculated by using Eqs.~\ref{eq20} and~\ref{eq33} at zero baryon chemical potential. Again, we take a constant value of hadronic cross-section ($\sigma$) equal to 11.3 mb while calculating the relaxation time ($\tau$). The upper panel represents $\zeta$ versus $T$ plot for different values of non-extensive parameter, $q$. We observe that $\zeta$ increases slightly with $T$ for a particular $q$, but as $q$ increases it starts saturating with the increasing $T$. In the middle panel, we show the variations of bulk viscosity to entropy density ratio ($\zeta/s$) at $\mu_B$ = 0 for various $q$-values. We find that $\zeta/s$ decreases rapidly with $T$ for all $q$. The value of $\zeta/s$ is lower for higher $q$-values. We also observe that $\zeta/s$ changes rapidly at a particular $T$, which is found to be lower for higher $q$-values. These findings suggest that the degree of approach towards an asymptotic value is sensitive to the non-extensive parameter, $q$. This goes inline with the expectations of dissipative properties of a system approaching thermal equilibrium. In the lower panel, we show the upper mass cutoff dependence of $\zeta/s$ for $q$ = 1.07. We observe a similar effect of mass cutoff on this ratio as observed in the case of $\eta/s$ $i. e.$ as we increase the upper mass cutoff of hadron resonance gas its magnitude decreases, which is due to increase of entropy density after adding the heavier resonances in the system.


\begin{figure}[h]
\centering
\includegraphics[width = 250px, height = 180px]{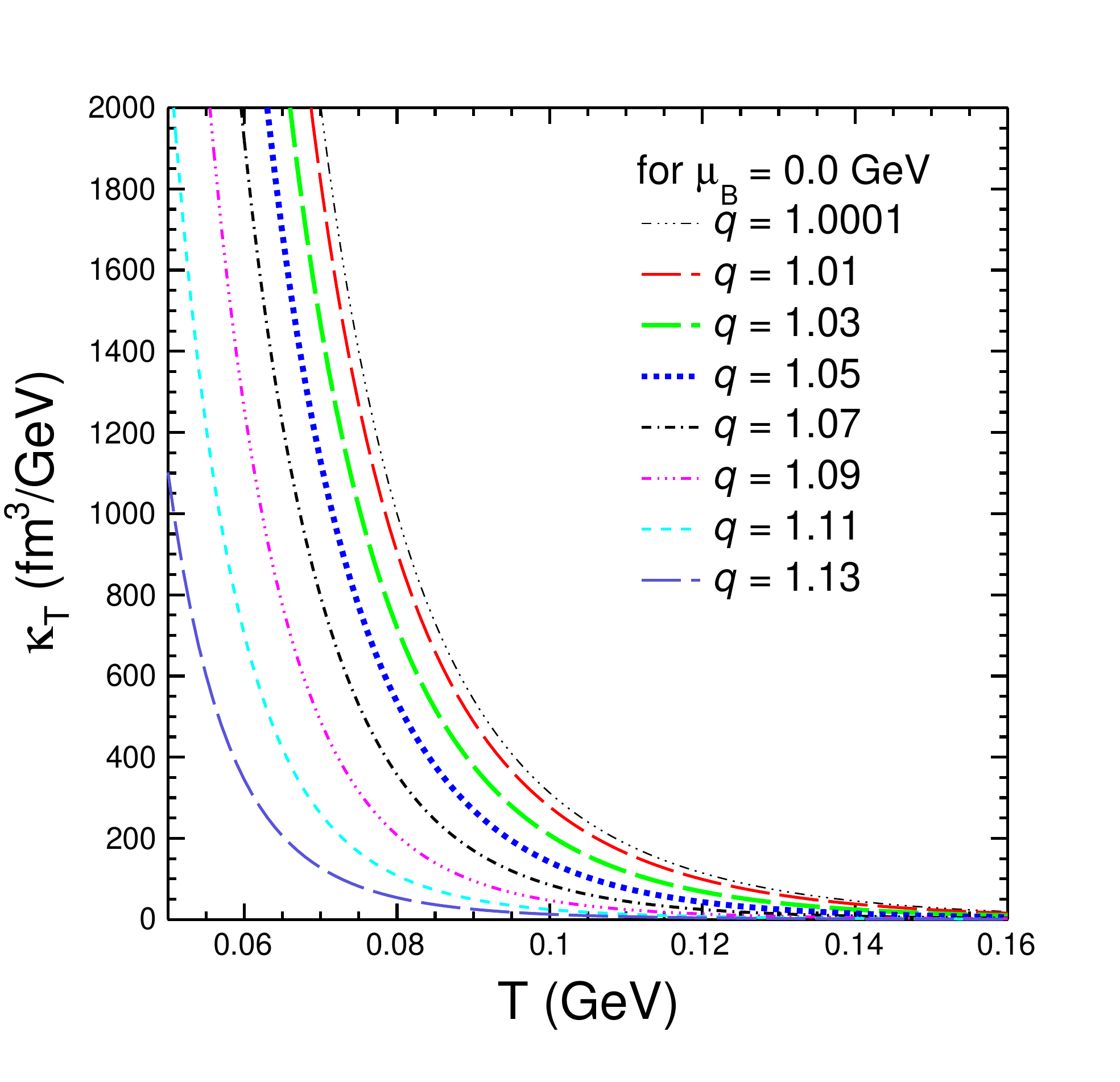}
\caption{(Color online) Isothermal compressibility as a function of temperature for different values of $q$, with $\mu_B$ = 0.}
 \label{kappa_mu0} 
\end{figure}

\begin{figure}[h]
\centering
\includegraphics[width = 250px, height = 180px]{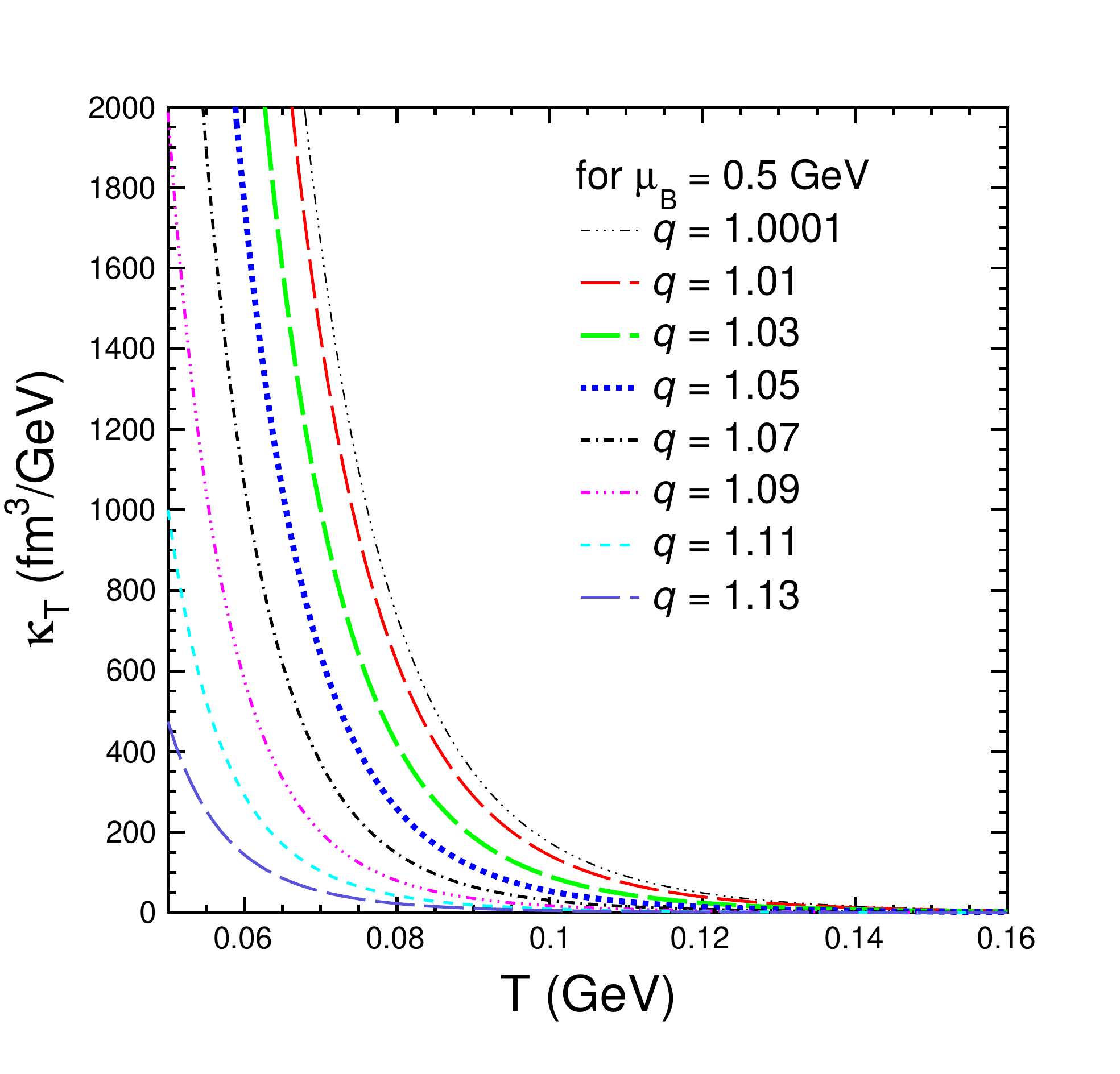}
\caption{(Color online) Isothermal compressibility as a function of temperature for different values of $q$, with $\mu_{B}$ = 0.5 GeV.}
 \label{kappa_mu5}
\end{figure}


\begin{figure}[h]
\centering
\includegraphics[width = 250px, height = 180px]{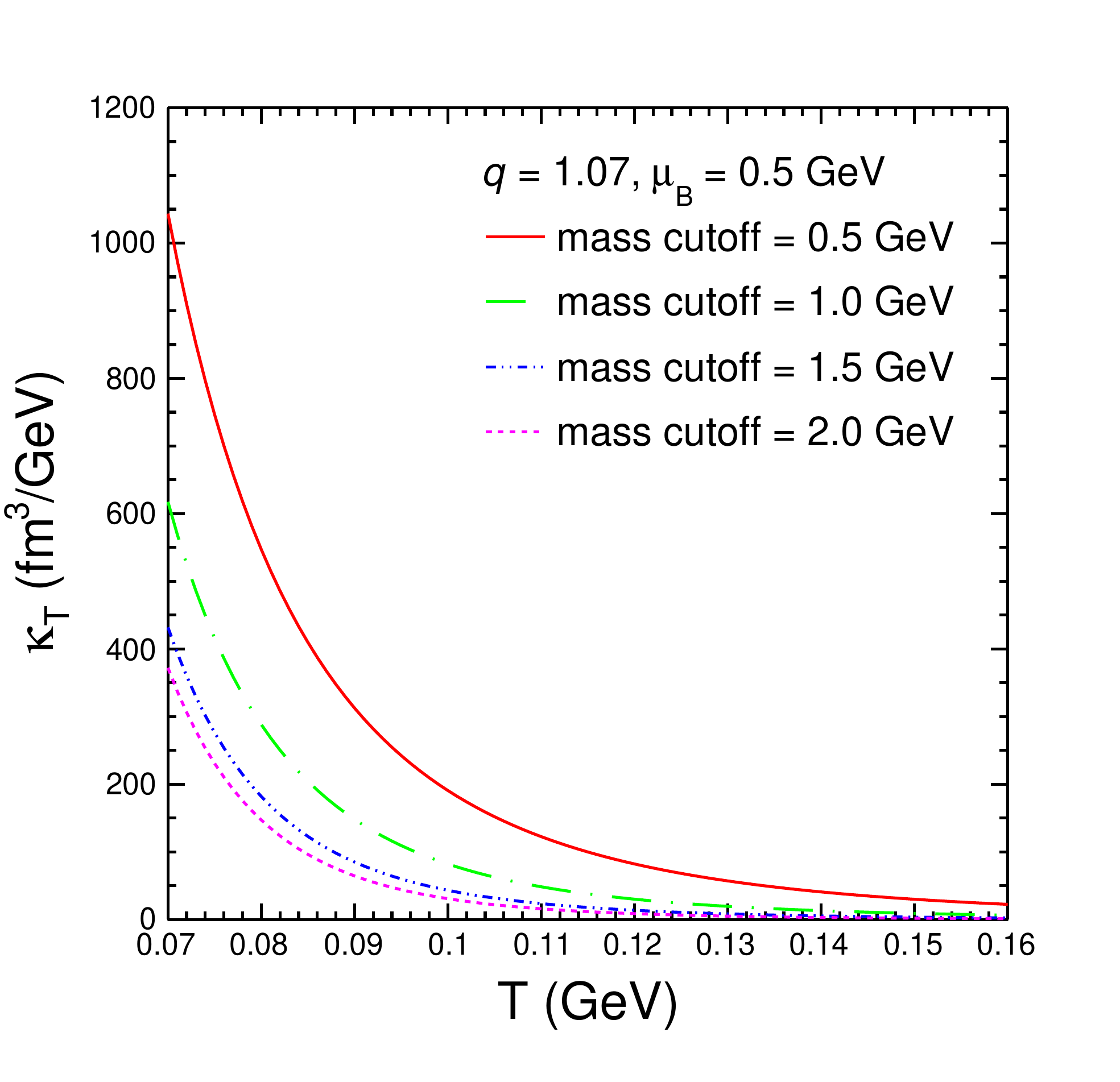}
\caption{(Color online) Isothermal compressibility as a function of temperature for different mass cutoff, with $\mu_B$ = 0.5 GeV and $q=1.07$}
 \label{kappa_mu5_107}
\end{figure}

\begin{figure}[h]
\centering
\includegraphics[width = 250px, height = 180px]{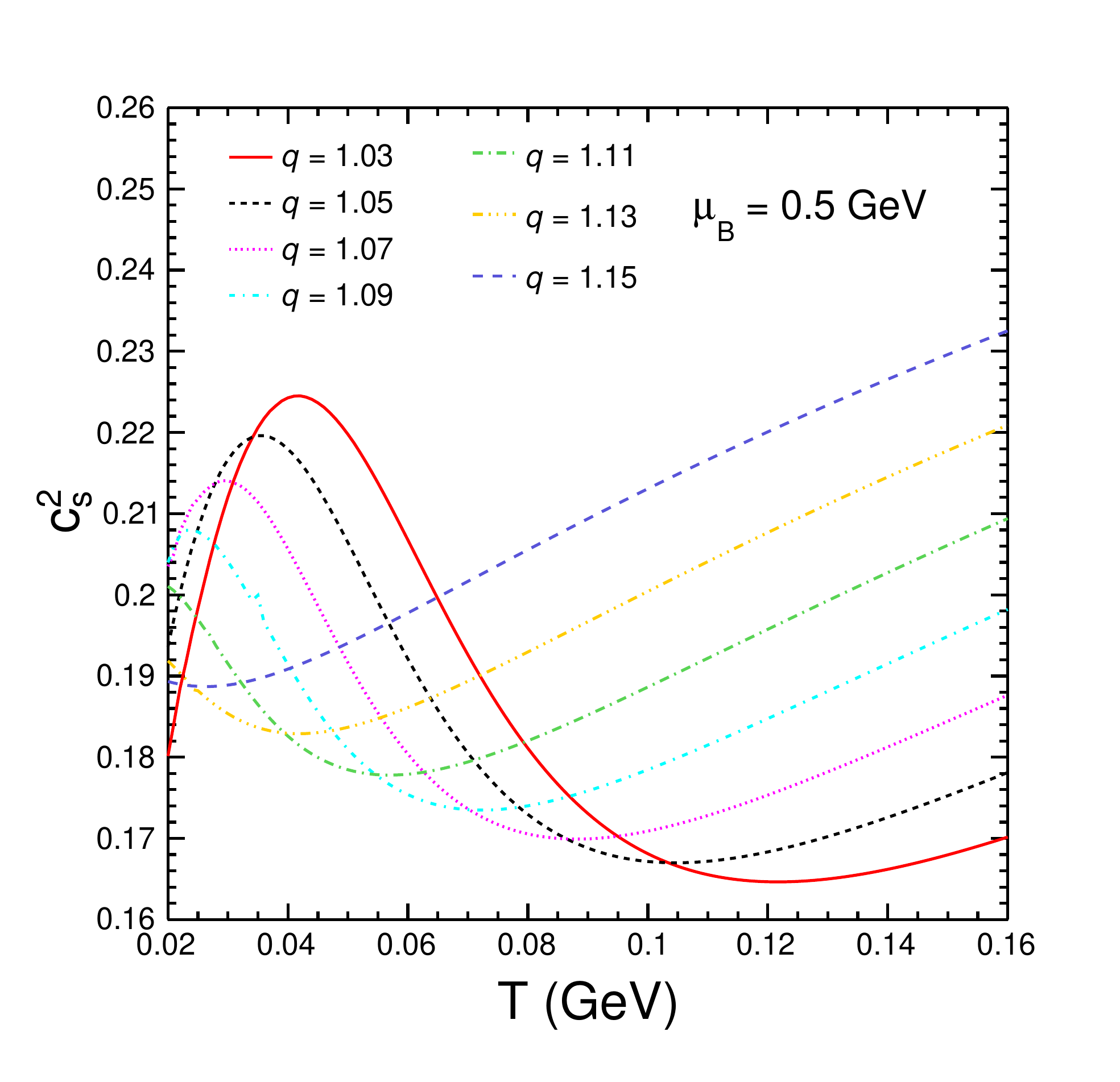}
\caption{(Color online) Speed of sound as a function of temperature for various values of $q$, with $\mu_B$ = 0.5 GeV.}
\label{cs}
\end{figure}

Next, we discuss one of the important thermodynamic observables $i. e.$ isothermal compressibility ($\kappa_T$), which has drawn considerable interest in these days. It is very interesting to quantify this observable for a hot and dense hadron gas formed in hadronic and heavy-ion collisions. In Ref.~\cite{Mukherjee:2017elm}, $\kappa_T$ is first time estimated for hadrons produced in heavy-ion collisions in extensive statistics using Hadron Resonance Gas (HRG) model. In this work, we have used non-extensive statistics for the calculation of $\kappa_T$. In Fig.~\ref{kappa_mu0}, results of $\kappa_T$ as a function of temperature at $\mu_B$ = 0 are shown. It is observed that $\kappa_T$ initially decreases rapidly at low temperatures and becomes constant at higher temperatures for all the $q$-values. The temperature at which $\kappa_T$ becomes constant depends on the $q$-value. As $q$ increases the value of $\kappa_T$ decrease and the temperature at which it becomes constant shifts towards the lower values, which suggests that there is a strong dependence of $\kappa_T$ on non-extensivity. Figure~\ref{kappa_mu5} describes the dependence of isothermal compressibility on temperature at $\mu_B$ = 0.5 GeV for various $q$. We find a similar behaviour of $\kappa_T$ in this case as observed for zero baryon chemical potential but comparatively has smaller values.     

Further, in order to see the effect of upper mass cutoff of hadrons on $\kappa_T$, we proceed as follows. We calculate $\kappa_T$ at $\mu_B$ = 0.5 GeV for $q$ = 1.07, which is shown in the Fig.~\ref{kappa_mu5_107} as a function of temperature. Again, a similar trend is obtained over all the temperatures as observed in the previous plots and as the upper mass cutoff increases, values of $\kappa_T$ decrease. This effect is more pronounced upto mass cutoff of 1 GeV and after that it gets weaker. $\kappa_T$ becomes independent of system temperature when higher resonances are added to the system. Hence a system is driven towards a critical behaviour, once we have resonances of higher masses as the ingredients of the system. 


\begin{figure}[h]
\centering
\includegraphics[width = 250px, height = 180px]{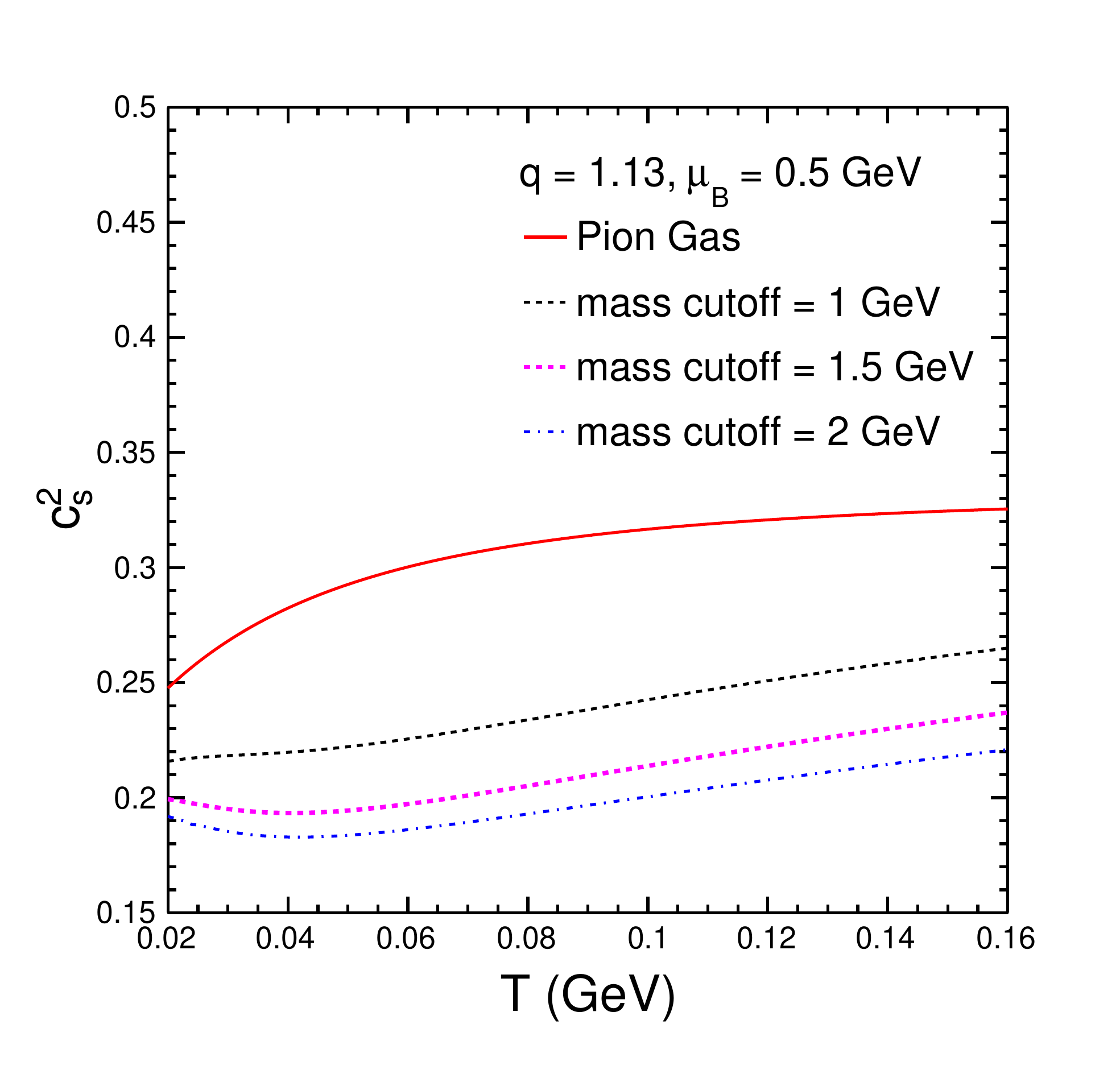}
\caption{(Color online) Speed of sound as a function of temperature for various mass cutoffs, at $\mu_B$ = 0.5 GeV.}
\label{cs_mass}
\end{figure}

Another important thermodynamical observable is speed of sound. We use Tsallis statistics to calculate the squared speed of sound ($c_s^2$) for finite baryon chemical potential. In Ref.~\cite{Khuntia:2016ikm}, the temperature and mass cutoff dependence of $c_s^2$ for various $q$-values have been studied in Tsallis statistics. It is noticed that the criticality in $c_s^2$, which is seen as the $q$-dependent peak in $c_s^2$, when studied as a function of temperature, disappears after $q$ = 1.13. We have shown $c_s^2$ as a function of temperature at $\mu_B$ = 0.5 GeV for different $q$-values in Fig.~\ref{cs}. We find that $c_s^2$ initially increases with $T$ and after reaching at a maximum value it starts decreasing. The maximum in this observable, which is related to the criticality shifts towards lower $T$ as $q$ increases and disappears after $q$ = 1.13. These findings are same as observed for the case of $\mu_B$ = 0 in Ref.~\cite{Khuntia:2016ikm}. 
In Fig.~\ref{cs_mass}, the mass cutoff dependence of $c_s^2$ is shown for $q$ = 1.13 and $\mu_B$ = 0.5 GeV. The criticality in $c_s^2$ disappears when lower mass cutoff is reached, which advocates that criticality of the system depends on the number of hadrons present in the system. Also, we find that $c_s^2$ decreases with the mass cutoff, which goes in line with the previous observations~\cite{Khuntia:2016ikm,Castorina:2009de}.              

\section{Summary}
\label{summary}
In summary, we have studied the transport coefficients using the relativistic non-extensive Boltzmann transport equation (NBTE), where we employ the relaxation time approximation for the collision integral. Here, we have extended the approach using Tsallis non-extensive statistics as $q$-equilibrium distribution function and calculated shear and bulk viscosity to the entropy density ratios. We have also calculated isothermal compressibility and squared speed of sound using Tsallis statistics at finite baryon chemical potential for different $q$. The important findings of this work are summarized as follows:
\begin{enumerate}

\item We have found that the ratios of shear and bulk viscosity to entropy density have a strong dependence on 
non-extensivity. Their values decrease with the increasing $q$-parameter with the temperature.

\item We have observed the effect of mass cutoff on these ratios and as mass cutoff increases the magnitudes of these ratios decrease. These effects are more pronounced at higher values of temperature.

\item We have first time studied the effect of non-extensive parameter, $q$ on isothermal compressibility. We have studied the temperature dependence of isothermal compressibility for various $q$-values and observed that it decreases rapidly with temperature and becomes constant at higher temperatures. The temperature at which the isothermal compressibility shows a saturation, may indicate an appearance of criticality in the system, which is dependent on $q$ and shifts towards lower values of temperatures for higher $q$-values.

\item While studying the temperature dependence of isothermal compressibility for various $q$-values, it has been found that the values of $\kappa_T$ are smaller for higher $q$. It has also been found that, its values decrease at finite baryon chemical potential in comparison to that obtained at zero baryon chemical potential.

\item We have found that, isothermal compressibility strongly depends on the upper mass cutoff of hadron resonances and decreases with the increasing mass cutoffs. It becomes independent of system temperature when higher resonances are added to it. Hence a system is driven towards a critical behaviour, once we have resonances of higher masses.

\item  It is observed that as the non-extensivity of the system increases the criticality observed as a peak in the squared speed of sound versus T plot shifts towards lower temperatures and disappears after $q$ = 1.13. We have also noticed a strong dependence of criticality on mass cutoff. As upper mass cutoff decreases, the criticality of the system disappears.   

\end{enumerate}

\section*{Acknowledgements}
NK and ST acknowledge the financial support by DST INSPIRE program of Government of India. SKT and RNS acknowledge the financial supports from ALICE Project No. SR/MF/PS-01/2014-IITI(G) of Department of Science $\&$ Technology, Government of India. The authors would like to thank Professor S. Gavin for useful discussions.

\end{document}